# Nonperfused Retinal Capillaries - A New Method Developed on OCT and OCTA

**Running head:** In vivo Imaging of Nonperfused Retinal Capillaries

Min Gao, MS,[1,2] Yukun Guo, MS,[1,2] Tristan T. Hormel, PhD,[2] Jie Wang, PhD,[2] Elizabeth White, MS,[2] Dong-Wouk Park, MD,[2] Thomas S. Hwang, MD,[2] Steven T. Bailey, MD,[2] Yali Jia, PhD[1,2,*]

[1]Department of Biomedical Engineering, Oregon Health & Science University, Portland, OR, USA

[2]Casey Eye Institute, Oregon Health & Science University, Portland, OR 97239, USA

**Corresponding author:** Yali Jia, PhD. Address: 515 SW Campus Dr, Portland, OR 97239. Phone number: 503 494 1053. Email: **jiaya@ohsu.edu**

**Abstract**

**Purpose**

To develop a new method to quantify nonperfused retinal capillaries (NPCs) by using co-registered optical coherence tomography (OCT) and OCT angiography (OCTA), and to evaluate NPCs in eyes with age-related macular degeneration (AMD) and diabetic retinopathy (DR).

**Design**

Retrospective, cross-sectional study.

**Methods**

In this study, a total of 153 participants, including 42 healthy controls, 78 with AMD, and 33 mild to moderate nonproliferative DR, were scanned with multiple consecutive 3×3-mm OCT/OCTA scans using a commercial device (Solix; Visionix/Optovue, Inc., California, USA). We averaged multiple registered OCT/OCTA scans to create high-definition volumes. The deep capillary plexus slab was carefully defined and segmented. A novel deep learning denoising algorithm removed tissue background noise from capillaries in the *en face* OCT/OCTA. The algorithm segmented NPCs by identifying capillaries from OCT images without corresponding flow signals in the OCTA. We then investigated the relationships between NPCs and known features in AMD and DR.

**Results**

The denoised *en face* OCT/OCTA clearly revealed the structure and flow of the capillaries. The automatically segmented NPC achieved an accuracy of 88.2% compared to manual grading of DR. Compared to healthy controls, both the mean number and total length (mm) of NPCs was significantly increased in eyes with AMD and eyes with DR (Mann-Whitney U test, $P < 0.001$, $P <0.001$). Compared to early and intermediate AMD, the number and total length of NPCs were significantly higher in advanced AMD (number: $P<0.001$, $P<0.001$; total length: $P = 0.002$, $P =0.003$). Geography atrophy, macular neovascularization, drusen volume, and extrafoveal avascular area (EAA) significantly correlated with increased NPCs ($P<0.05$). In eyes with DR, NPCs correlated with the number of microaneurysms and EAA ($P<0.05$). The presence of fluid did not significantly correlate with NPCs in AMD and DR.

**Conclusion**

A deep learning-based algorithm can segment and quantify retinal capillaries that lack flow using colocalized OCT/OCTA. This novel biomarker may be useful in AMD and DR.

**Introduction**

Optical coherence tomography (OCT) is a noninvasive imaging technique that captures detailed cross-sectional images of retina.[1] Expanding on this technology, OCT angiography (OCTA) provides functional assessment by visualizing blood flow (functional circulation) within the retinal microvasculature.[2] A single OCTA scan produces co-registered OCT and OCTA volumes, enabling the simultaneous extraction of both retinal structural and angiographic features. Retinal fluid and microaneurysms on OCT,[3,4] along with retinal avascular areas on OCTA,[5–8] are typical lesions of diabetic retinopathy (DR). Geography atrophy (GA) and drusen on OCT,[9–13] macular neovascularization (MNV) on OCTA[14] are characteristic features of age-related macular degeneration (AMD). The accurate identification and segmentation of these features from OCT/OCTA can potentially stratify DR and AMD severity and predict the progression of a variety of eye diseases, informing clinical management strategies.[15–18]

In diabetic retinopathy, sections of capillaries lose pericytes and endothelium, becoming acellular basement membrane tubes as seen in histopathologic studies. These tubules do not have lumens and correlate to areas of nonperfusion on angiography.[19–21] Combining OCT and OCTA signals allows *in vivo* quantification of these occluded capillaries. This approach has advantages over previous approaches to measure ischemia such as measurement of avascular areas, vessel density, or fractal dimensions. There are variations in normal intercapillary space, particularly around the foveal avascular zone (FAZ), where nonperfusion is most consequential but may not be helpful in distinguishing disease from control due to this variation. This approach also avoids the false detection of ischemia caused by low signal artifacts due to media opacities or low signal.[6]

There are additional pathologic mechanisms that cause capillaries to have reduced perfusion. In diabetes, an increase in diacylglycerol concentration activates various forms of protein kinase, which sensitize the contractile mechanism of the retinal arteriole, resulting in hypoperfusion. This may be an important pathophysiologic change, especially in early stage of the disease.[22]

In age-related macular degeneration, recent OCTA-based studies have reported changes in retinal capillary perfusion in both dry AMD[9] and wet AMD.[23] While it is primarily a disease of the outer retina and the choroid, decreased perfusion in the inner retina and deep vascular plexus may result from photoreceptor degeneration.[9,23,24] It is also possible that the retinal vasculature plays a role in the progression of disease.[9,23,25]

A previous study employed adaptive optics scanning light ophthalmoscope (AOSLO) infrared structural and fluorescein angiographic paired images to detect NPCs.[26] However, AOSLO limited the study to the fovea only and NPCs were demonstrated in healthy controls through visual comparison side by side. AOSLO is not a volumetric imaging modality and so cannot inform our understanding of NPC distribution within the various retinal tissue layers. Lynch et al. visualized NPCs by comparing contrast-inverted *en face* OCT and OCTA, generated by ten-averaged 3×3-mm scans centered at the fovea.[27] The *en face* OCT provided structural capillary information while the *en face* OCTA displayed flow signal within them. By comparing these two modalities, NPC segments in DR at the innermost border of the FAZ were identified. Despite this earlier observation, the automatic identification and segmentation of capillary structures in *en face* OCT remains challenging due to the low contrast inherent in these images.

Deep learning has proven to be a powerful method for extracting features from biomedical images for super-resolution reconstruction,[28–30] lesion detection, and segmentation.[11,31–34] In this study, we designed a deep learning-based method to denoise *en face* OCT/OCTA and automatically segment NPCs from paired OCT and OCTA information. The extent and distribution of NPCs were evaluated for healthy eyes and eyes with AMD and DR. We also explored the relationship between NPCs and known AMD and DR biomarkers.

## Methods

**Data acquisition**

The Institutional Review Board of Oregon Health & Science University approved this study. Informed consent was obtained from all participants, and the study adhered to the Declaration of Helsinki. The inclusion criteria for the control group required participants to have no history of retinal diseases, hypertension, or diabetes. AMD participants could not have other retinal diseases or previous intraocular surgery, except for cataract surgery. Patients with hypertension were included in the AMD group, as hypertension is common in the elderly and may coexist with AMD. For the DR groups, the inclusion criteria required participants to have a confirmed diagnosis of DR, ranging from mild to moderate nonproliferative DR (NPDR), based on clinical examination. Participants were excluded if they had a history of any other macular diseases, such as epiretinal membrane or vitreomacular traction syndrome.

Six 3×3-mm macular OCT/OCTA scans with a 400×400-pixel transverse sampling density from one and only one eye of each participant were obtained continuously within 5 minutes using a commercial 120-kHz spectral-domain OCT system (Solix; Visionix/Optovue, Inc., California, USA). The split-spectrum amplitude-decorrelation angiography algorithm implemented in this instrument was used to generate the OCTA data.[2] Twelve of the orthogonal scans (x-fast and y-fast) were registered to generate six motion-free volumes.[35] These six volumes were registered and merged to obtain a high-definition OCTA volume. A projection-resolved OCTA algorithm was applied to suppress projection artifacts throughout the entire volume.[36,37] The internal limiting membrane, inner plexiform layer, inner nuclear layer (INL), outer plexiform layer (OPL), and Bruch's membrane were segmented by a guided bidirectional graph search algorithm.[38] Expert graders manually corrected errors on these layers' segmentation.

**Multiple three-dimensional OCT/OCTA volumetric registration**

We developed a volumetric registration algorithm to register and average six OCTA volumes to generate a high-definition OCTA volume. In this algorithm, we first generated *en face* OCTA from all the repeated scans. The *en face* OCTA with the highest signal strength index (SSI) in all repeated scans served as the fixed image (Figure 1A) to which the other scans were registered (Figure 1B). Then, an intensity-based automated image registration algorithm was used to register these moving images with the fixed image[38,39]. The registration algorithm can produce transformation matrixes, which contain translation, rotation, and scaling parameters (Figure 1C). These transformation matrixes were applied to the whole OCT/OCTA volumes for the scans to be registered (Figure 1D). Thus, each pair of C-scans between fixed and all moving scans were separately registered. Then, a normalized correlation method was deployed on each pair of A-lines in the fixed and moving cross-sections (Figure 1E) to generate the registered volumes (Figure 1F). Finally, these registered OCT/OCTA volumes were averaged to produce a high-definition volume (Figure 1G).

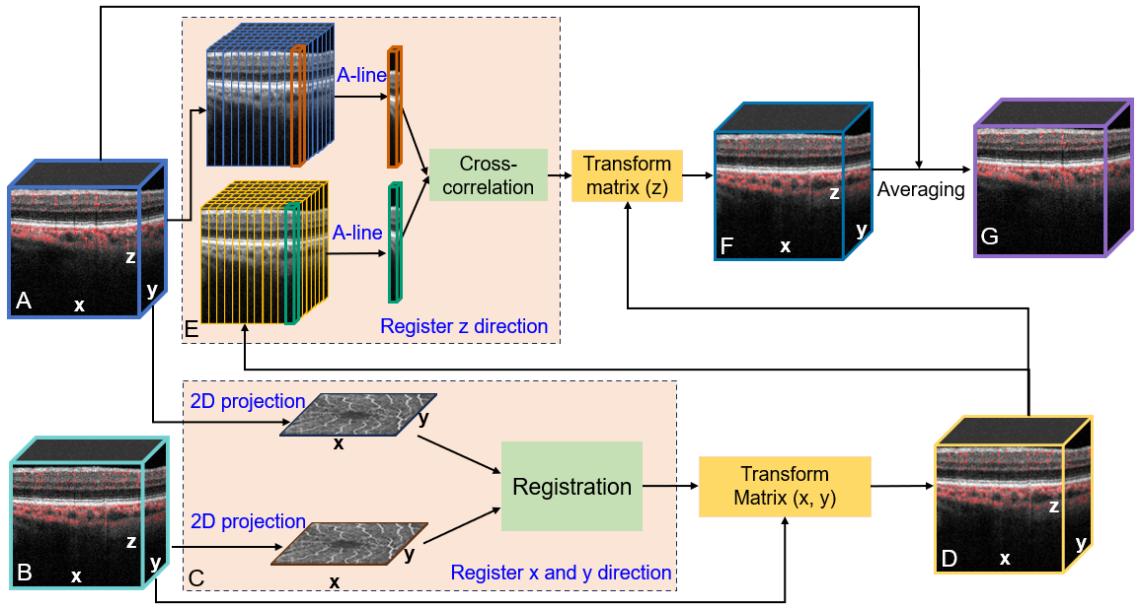

**Figure 1.** (A) Fixed OCTA volume. (B) Moving OCTA volume. (C) Registering moving angiogram with fixed angiogram and applying transform matrix to OCT/OCTA volumes. (D) Moving OCT/OCTA volume registered in the x- and y-directions. (E) Registering each pair of A lines in OCT volumes and applying the transform matrix to OCT/OCTA volumes. (F) Moving OCT/OCTA volumes registered in the z-direction. (G) Averaging the fixed and registered moving OCT/OCTA volumes to generate high-definition OCT/OCTA volumes.

**Projection of *en face* OCT/OCTA**

Intermediate capillary plexus (ICP) angiograms were projected in a slab between the outer 20% of the ganglion cell complex and the inner 50% of the INL. We used redefined boundaries of deep capillary plexus (DCP) to project *en face* OCT/OCTA (Figure 2). *En face* OCT/OCTA of DCP were projected in a slab 9 μm above the lower boundary of the INL and 12 μm below the lower boundary of the INL. *En face* OCT and OCTA of ICP and DCP were produced by mean and maximum projection of the OCT and OCTA signal,[40] respectively. In the FAZ, the *en face* OCT of DCP may contain hyperreflective spots (arrowhead position in Figure 2B), which can greatly reduce the image contrast; moreover, pupil vignetting also can reduce the quality of *en face* OCT. Thus, a 0.6 mm radius circle centered over the fovea was excluded. An annular zone surrounded by a concentric ring 0.6 to 1.5 mm from the fovea was used to characterize deep capillaries (Figure 2).

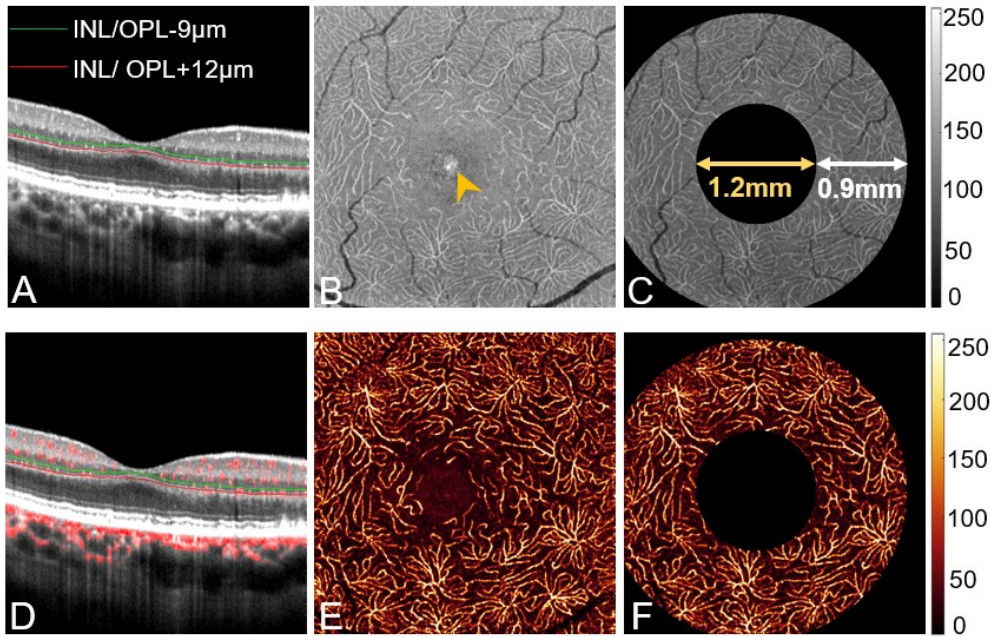

**Figure 2.** OCT/OCTA images of deep capillaries. (A) The deep capillary plexus (DCP) on cross-sectional OCT. The upper (green curve) and lower (red curve) boundaries of the region containing capillaries in DCP are indicated. (B) *En face* OCT of DCP using mean projection. (C) An annular zone surrounded by concentric rings 0.6mm to 1.5mm from the fovea was used to characterize capillaries due to strong noise and low contrast. (D) OCTA signal overlaid on cross-sectional OCT. (E) *En face* OCTA of DCP by maximum projection. (F) An annular zone of the OCTA signal was cropped for analysis.

*En face* **OCT/OCTA denoising**

NPCs can be detected by comparing the OCT/OCTA images. However, the reflectance of capillaries and surrounding tissue significantly reduces the contrast in *en face* OCT, making it challenging to differentiate capillaries from tissue. In this study, we used *en face* OCTA, which have relatively less noise, to train a deep learning denoising model and applied this model to *en face* OCT.[41] We used the denoised *en face* OCT/OCTA to generate the binary *en face* OCT/OCTA to segment NPCs.

**An end-to-end network architecture**

We designed an end-to-end convolutional neural network (CNN) to generate background noise-resolved and enhanced *en face* OCTA of the intermediate and deep retinal slabs. The OCT angiograms (Figure 3A) with different noise levels were input into the network to train a model. The ground truth was the reconstructed noise-free and high-definition angiograms generated from DCARnet, our previously validated denoising network.[29] The current network comprises a pyramid network (Figure 3B) with dense block (Figure 3C) and selective kernel subnetworks (SKNet) (Figure 3D).[42] The pyramid network can provide multi-scale information using a hierarchical architecture. The features from the different scales were extracted by the dense block, which encourages the network to reuse features learned by earlier layers in the subsequent layers. This can be highly beneficial for feature extraction and learning representations, as it helps combat the vanishing gradient problem and promotes feature reuse. SKNet can improve the ability of CNN to capture multi-scale information by selectively integrating kernels with different receptive field sizes and resolutions, allowing the network to extract both local and global information in the input. Finally, the selected useful features from the different scales were fused to the output layer to generate noise-free *en face* OCTA (Figure 3E). The ReLU activation function was used in each convolutional layer, except for the last convolution output layer.

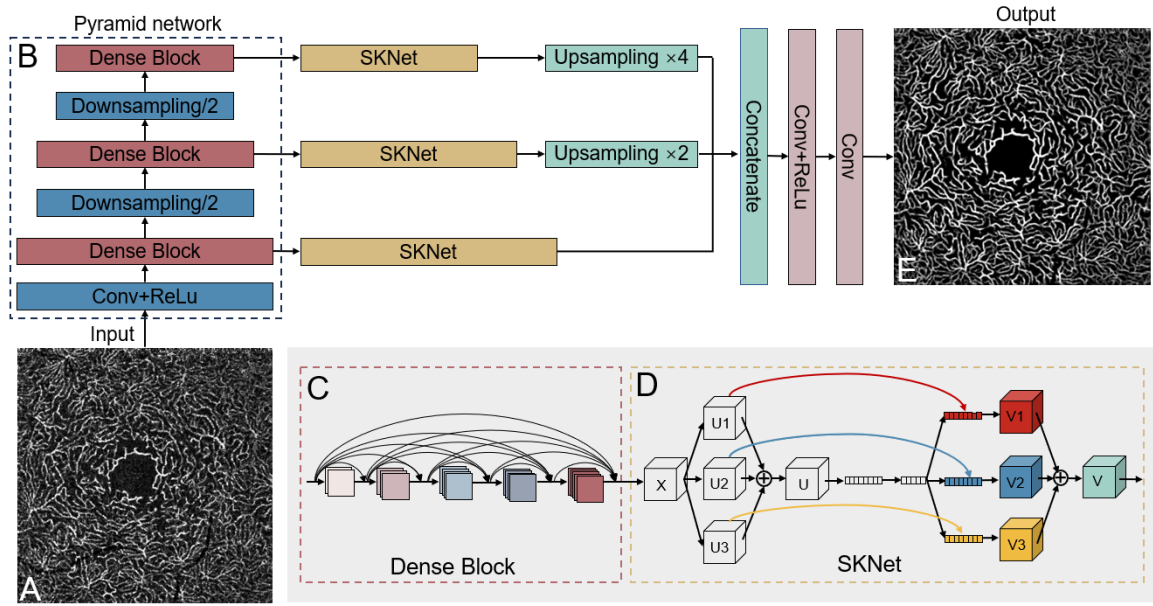

**Figure 3.** Network architecture for *en face* OCTA denoising and enhancement. (A) The input is *en face* OCTA with different noise intensities. The input was downsampled a different number of times using a pyramid structure (B), and features were extracted by a dense block (C) to pass to a selective kernel network (SKNet) (D). SKNet fused features from the effective receptive fields of different kernel sizes using softmax attention in each scale. The effective features in each scale were selected and concatenated to output noise-free *en face* OCTA (E).

**Subjects and ground truth generation for *en face* OCTA denoising**

Each eye was repeatedly scanned using a 3×3-mm scan pattern centered on the macula (Figure 4A). Among these multiple repeated scans, those exhibiting high motion artifacts and those that were off-center were excluded. Many studies have also shown that averaging of multiple *en face* OCTA can enhance image quality.[43,44] Registered and averaged OCT/OCTA volumes significantly suppressed the noise and improved the contrast and vascular connectivity (Figure 4B). We selected 36 scans from 78 averaged scans from 78 eyes with AMD, ensuring low noise and high quality, to serve as the training dataset. In a previous study, we proposed DCARnet to reconstruct relatively low-resolution 6×6-mm *en face* OCTA using high-resolution 3×3-mm *en face* OCTA. This method does not generate false signals when the noise intensity is not extremely high.[29] Therefore, we applied DCARnet (Figure 4C) to these 36 scans to produce noise-free *en face* OCTA, which were used as the ground truth (Figure 4D).

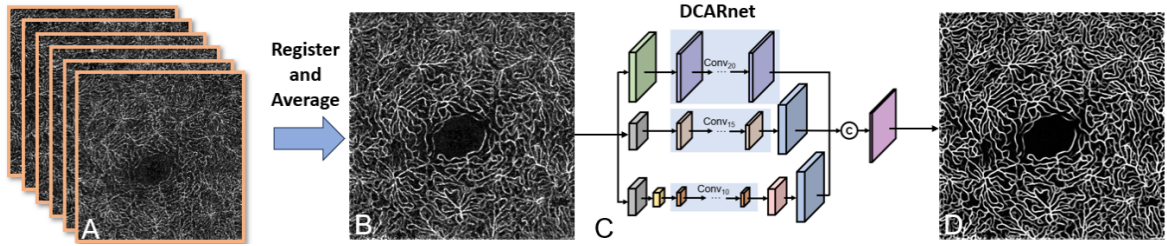

**Figure 4.** Multiple single OCT/OCTA scans (A) were registered and averaged to generate an OCT/OCTA volume (B) *En face* OCTA of ICP and DCP were delivered to DCARnet. (C) to reconstruct a noise-free background image for an enhanced ground truth (D).

**Training parameters**

The angiograms were reconstructed by DCARnet to generate the ground truth using 36 scans, which were registered and averaged from a total of 257 scans from 36 eyes. To enhance the ability of our network to remove strong noise from angiograms, we obtained the noise in the *en face* OCTA by subtracting the ground truth from the original angiograms of single and averaged scans. We then added 0 to 0.5 times of the real noise, in intervals of 0.1, to the original single and averaged angiograms to generate angiograms with varying noise intensities as network inputs. We also applied a flat-field correction to the original angiograms to reduce shading distortion and darkening. The correction uses Gaussian smoothing with a standard deviation of 30 to approximate the shading component of the angiogram. The illumination-compensated *en face* OCTA were also input to train the network.

To further augment our training data, we also included ICP angiograms as input, given the similarity between ICP and DCP features. The dataset was also augmented through horizontal and vertical flipping, random rotation and transposition, and random cropping. This resulted in the generation of 4102 ICP and DCP angiograms with different noise intensities. The training dataset comprised 3276 ICP and DCP angiograms, while the validation dataset consisted of 826 ICP and DCP angiograms.

The size of input images was 320×320 pixels and the batch size was set to 4. We trained the network using the Adam optimizer with an initial learning rate of 0.0001 and a global learning rate decay strategy. Training was conducted on a GPU server equipped with four NVIDIA GeForce RTX 3090 graphics cards, utilizing two of these cards for this study. The network was trained using a linear combined loss function incorporating mean square error and structural similarity.

**Applying the OCTA denoising model to OCT**

To resolve background noise of the original *en face* OCT (Figure 5A), we first adjusted the grayscale histogram to match that of the *en face* OCTA. This adjustment enhanced the contrast of the *en face* OCT, producing capillary features similar to those in the *en face* OCTA (Figure 5B). Thus, we were able to apply the deep learning OCTA denoising model to the *en face* OCT (Figure 5C, 5D). The deep learning model output noise-free *en face* OCT/OCTA, which were then binarized. Nonperfused retinal capillaries (NPCs) were generated by subtracting the flow signal identified by OCTA from the structural capillaries detected by OCT.

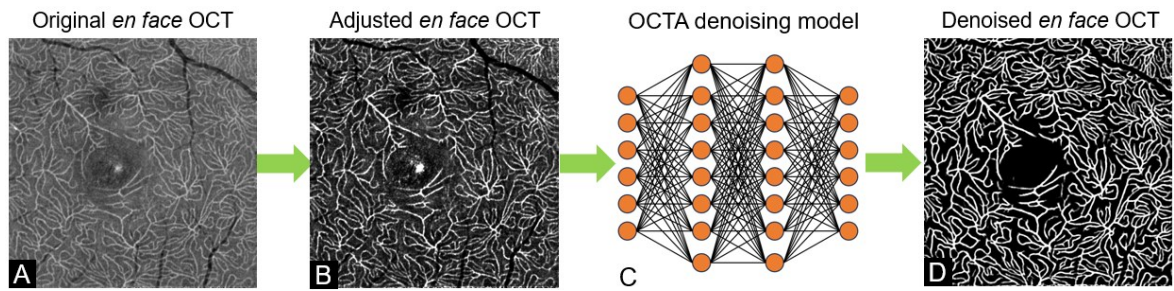

**Figure 5.** (A) Original *en face* OCT with strong background noise and low contrast. (B) Adjusted *en face* OCT exhibits similar features to those of capillaries on *en face* OCTA. (C) Trained *en face* OCTA denoising model. (D) Background noise resolved *en face* OCT.

**Evaluation and characteristics of nonperfused retinal capillaries**

We manually graded the accuracy of NPC segmentation (Y.G., M.G.) by overlaying the segmented NPCs on the *en face* OCT/OCTA and checking if the segmented NPCs were real capillaries on *en face* OCT and if there was flow in the segmented NPCs on *en face* OCTA. We also checked the reflectance and flow signal of the segmented NPCs on cross-sectional OCT/OCTA.

We calculated the number and length of NPCs after skeletonization and compared them in healthy eyes and eyes with diseases, including AMD and DR, excluding nonperfused segments less than 75 μm in length. We specifically investigated whether significant differences existed in NPCs between varying severities of AMD. We explored the relationship between the number and length of NPCs and known features of AMD or DR, including GA, MNV, retinal fluid, drusen, extrafoveal avascular area (EAA), and microaneurysm. Retinal fluid, drusen, EAA, and GA were automatically segmented using deep learning-based methods.[3,5,10,13] The presence of MNV, MA was graded by ophthalmologists (T.H., S.B.).

**Statistical analysis**

We first performed a Kolmogorov-Smirnov test to check if measurements were normally distributed. Then a Mann-Whitney U test was applied to test if there is a difference between two groups at a 5% significance level. For comparisons involving more than two groups, we used the Kruskal-Wallis test to assess whether at least one pair of groups showed significant differences. When significant differences were identified, Dunn's post hoc tests with Bonferroni correction were performed to determine which specific group pairs had statistically significant differences. Pearson correlation coefficient was calculated to test the relationship between two continuous variables. Univariate logistic regression was used to suggest the relationship between the presence of a biomarker and the number or length of NPCs.

**Results**

**Clinical characteristics of study participants**

The clinical characteristics of study participants included healthy controls (N= 43), patients with AMD (N = 78), and those with mild to moderate NPDR (N = 33) (Table 1). The mean ages for the groups were 45.9 ± 17.5 years for the healthy controls, 77.1 ± 7.5 years for the AMD group, and 64.1 ± 11.6 years for the mild to moderate NPDR group. The number and length of NPCs did not show dependency on age in healthy controls. Compared to healthy controls, both the number and length of NPCs significantly increased in eyes with AMD and mild to moderate NPDR (P < 0.001, P < 0.001, Mann-Whitney U test).

**Table 1. Clinical characteristics of study participants with AMD, DR and controls.**

| Parameters | Healthy controls (N=43) | AMD (N=78) | Mild to moderate NPDR (N=33) |
|---|---|---|---|
| Age (mean±standard deviation) | 45.9 ±17.5 | 77.1±7.5 | 64.1±11.6 |
| Sex (Male/Female) | 22/21 | 31/47 | 19/14 |
| Hypertension (Yes/No) | 0/43 | 43/35 | 20/13 |
| Diabetes (Yes/No) | 0/43 | 9/69 | 33/0 |
| Number of NPCs | 1.79 ± 1.42 | 17.50 ± 13.12 | 19.82 ± 8.95 |
| Total length of NPCs (mm) | 0.35 ± 0.40 | 3.30 ± 3.34 | 5.63 ± 2.98 |

AMD: age-related macular degeneration, NPDR: nonproliferative diabetic retinopathy, NPCs: nonperfused retinal capillaries.

**Performance of nonperfused retinal capillaries segmentation**

Our method segmented 654 NPCs in 33 eyes with mild to moderate NPDR. However, due to the presence of very strong noise with high signal intensity and structures resembling capillaries in *en face* OCT, 77 segments were erroneously detected as NPCs. The accuracy of NPC segmentation was 88.2% compared to manual graders. The segmented NPCs were further validated by examining the reflectance and flow signal on cross-sectional OCT/OCTA.

**Relationship between nonperfused retinal capillaries and known biomarkers in eyes with age-related macular degeneration**

Nonperfused retinal capillaries in 78 eyes with AMD (32 early AMD, 16 intermediate AMD, and 30 advanced AMD) were automatically segmented. There were no significant differences in the number and total length between early and intermediate AMD [number: 11.12±0.83 vs. 12.44±1.22, P = 0.492; total length (mm): 1.81±1.83 vs. 2.13±0.26, P = 0.428, post-hoc test with Bonferroni correction]. However, compared to eyes with early AMD, both the number and total length significantly increased in eyes with advanced AMD [number: 11.12±0.83 vs. 26.27±3.26, P <0.001; total length (mm): 1.81±1.83 vs. 2.13±5.39, P <0.001]. The number and total length of NPCs was also significantly higher in advanced AMD compared to intermediate AMD [number: 12.44±1.22 vs. 26.27±3.26, P =0.002; total length (mm): 2.13±0.26 vs. 2.13±5.39, P =0.003]. The presence of MNV (N=14) and GA (N=21) (Figure 6A) was associated with both the number [odds ratio (OR) = 1.227, P < 0.001; OR = 1.070, P = 0.016, univariate logistic regression] and the total length (OR = 1.624, P = 0.001; OR = 1.172, P = 0.017, univariate logistic regression) of NPCs (Table 2), with most NPCs situated within the GA region. The number and total length of NPCs showed no significant difference between eyes with only GA (N = 9) and those with only MNV (N = 16) [P=0.388, P = 0.329, Mann-Whitney U test]. EAA and drusen volume were related to the number (R = 0.554, P < 0.001; R = 0.254, P = 0.025, Pearson correlation) and total length (R = 0.566, P < 0.001; R = 0.237, P = 0.037, Pearson correlation) of NPCs (Figure 6B, 6D) and the majority of NPCs were located within avascular areas (Figure 6B). However, the presence of retinal fluid (N=19) was not related to either the number (P = 0.70, OR = 1.01, univariate logistic regression) or the total length (P = 0.46, OR = 1.05, univariate logistic regression) of NPCs (Figure 6C).

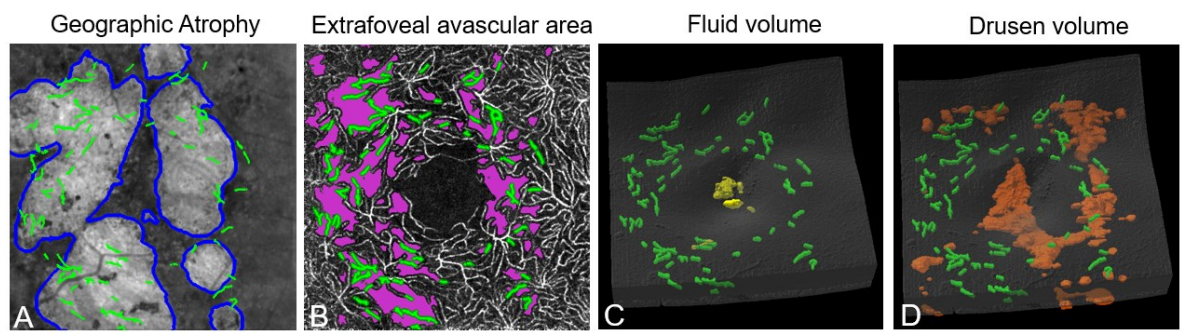

**Figure 6.** Nonperfused retinal capillaries (NPCs) and know biomarkers in an eye with advanced age-related macular degeneration. (A-D) Geographic atrophy (GA, blue contour), extrafoveal avascular area (EAA, violet), fluid (yellow) and drusen (orange) volume were automatically segmented using deep learning-based methods. The presence of GA, NPA and drusen volume significantly correlated with the number and total length of NPCs; however, the presence of retinal fluid was not related to either the number or total length of NPCs.

**Table 2. Relationship between nonperfused retinal capillaries and known biomarkers in eyes with age-related macular degeneration (N=78).**

|  | MNV presence (N=14) | GA presence (N=26) | Fluid presence (N=19) | EAA (N=78) | Drusen volume (N=78) |
|---|---|---|---|---|---|
| Number of NPC | *P=0.02, OR=1.07 | *P<0.001, OR=1.23 | *P=0.70, OR=1.01 | **P<0.001, R=0.55 | **P=0.02, R=0.27 |
| Total length of NPC (mm) | *P=0.02, OR=1.17 | *P=0.001, OR=1.62 | *P=0.46, OR=1.05 | **P<0.001, R=0.57 | **P=0.02, R=0.26 |

NPC: Nonperfused retinal capillary, MNV: Macular neovascularization, GA: Geographic atrophy, EAA: Extrafoveal avascular area. * Univariate logistic regression; **Pearson correlation. Correlation is significant at 0.05.

**Relationship between nonperfused retinal capillaries and known biomarkers in eyes with mild to moderate non-proliferative diabetic retinopathy**

We segmented NPCs in 33 eyes with mild to moderate NPDR. The results showed that the number of microaneurysms were associated with the number (P=0.03, R=0.37, Pearson correlation) and total length (P=0.01, R=0.46, Pearson correlation) of NPCs (Table 3). NPCs were frequently located adjacent to microaneurysms (Figure 7C). EAA was also correlated with the number (P<0.001, R=0.72, Pearson correlation) and total length (P<0.001, R=0.57, Pearson correlation) of NPCs. Most NPCs were located within avascular areas (Figure 7A, 7B). However, the presence of fluid (18/15) was unrelated to the number (P=0.10, OR=1.07, Univariate logistic regression) and total length (P=0.08, OR=1.27, Univariate logistic regression) of NPCs (Figure 7D).

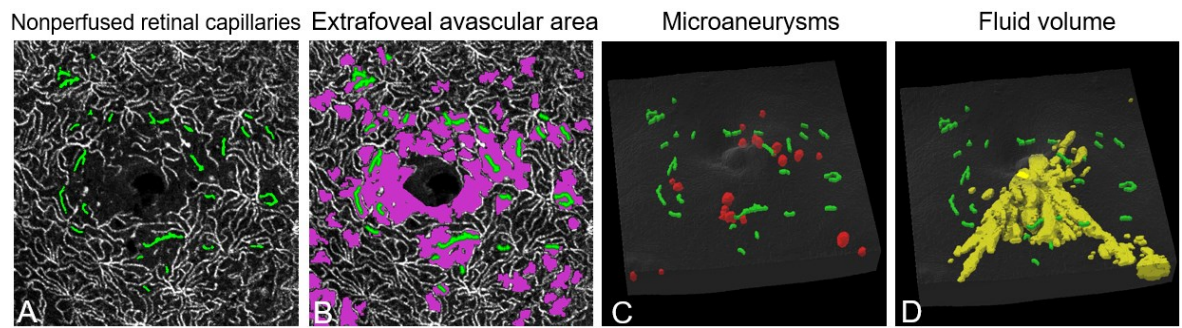

**Figure 7.** Nonperfused retinal capillaries (NPCs) and know biomarkers in an eye with moderate nonproliferative diabetic retinopathy. (A-D) NPCs (green), extrafoveal avascular area (EAA, violet), microaneurysms (red) and fluid (yellow) volume were automatically segmented using deep learning-based methods. The number of microaneurysms and NPA significantly correlated with the number and total length of NPCs; however, the presence of retinal fluid was not related to either the number or total length of NPCs.

**Table 3.** Relationship between nonperfused retinal capillaries and known biomarkers in eyes with diabetic retinopathy (N= 33).

|  | **Fluid presence (N=18)** | **Number of microaneurysm** | **Extrafoveal avascular area** |
| --- | --- | --- | --- |
| Number of NPC | *P=0.10, OR=1.07 | **P=0.03, R=0.37 | **P<0.001, R=0.72 |
| Total length of NPC (mm) | *P=0.08, OR=1.27 | **P=0.01, R=0.46 | **P<0.001, R=0.57 |

NPC: Nonperfused retinal capillary. * Univariate logistic regression. **Pearson correlation. Correlation is significant at 0.05.

## Discussion

In this study, we utilized co-registered OCT and OCTA to develop a novel method for the automatic segmentation and characterization of NPCs, leveraging deep learning techniques to enhance the accuracy and reliability of NPC detection. The number and length of NPCs are increased in various levels of AMD and DR, suggesting their potential as a valuable biomarker of disease.

It is straightforward to understand that one modality can reveal capillary structure (whether perfused or not), while another modality displays the blood flow within these capillaries. By fusing these images, we can effectively distinguish nonperfused (occluded) segments from perfused (non-occluded) ones. Traditionally, visualizing capillary structures has required high-resolution modalities equipped with adaptive optics (AO), such as AO-SLO or AO-OCT, for *in* *vivo* imaging. Although prior studies have demonstrated the use of commercial OCT for visualizing the innermost capillary loops at the FAZ border,[19] this visualization was limited to a small cohort of DR cases without the interference of edema. This highlights the need for advanced techniques to extend beyond the FAZ loop and address the signal attenuation challenges posed by pathologies, such as retinal edema.

A significant challenge in visualizing capillaries beyond the FAZ loop with OCT is differentiating true capillary structures from background noise. Given that a single capillary is only 5-10 microns wide—smaller than the transverse resolution of commercial OCT—its reflectance signal can be obscured by the surrounding retinal tissues. Our approach overcomes this by implementing four key steps.

First, we enhanced image resolution and contrast by improving signal-to-noise ratio using advanced three-dimensional registration and averaging techniques. Second, we carefully defined the study slab to further enhance the contrast between capillaries and the strong tissue background noise. Third, we developed a deep learning-based algorithm to denoise angiograms and applied it to eliminate the background tissue noise from *en face* OCT, revealing clear capillary structure and flow signal. Fourth, we utilized a multi-modal fusion of OCT and OCTA images to combine structural and functional information, enabling precise identification of NPCs. These innovations collectively enhanced our ability to reveal capillary vasculatures, achieving a segmentation accuracy of 88.2%.

This approach was validated in cohorts with AMD and DR at Oregon Health and Science University, revealing significant correlations between NPCs and known biomarkers. Key findings from our results include:
- Compared to early and intermediate AMD, both the number and total length of NPCs were significantly increased in advanced AMD. These findings indicate a progression in the extent of NPCs as AMD advances to more severe stages.

- In both AMD and DR, the majority of NPCs were located within avascular areas (Figure 6, Figure 7). This observation confirms our detection accuracy, as clusters of NPC segments form avascular regions.
- In AMD, the presence of GA was highly associated with NPCs, with most NPCs situated within the GA region (Figure 6). This finding aligns with previous research indicating reduced retinal vessel density in the deep capillary plexus and significant loss within GA regions.[9]
- In exudative AMD, we found a significant association between MNV and NPC. This is consistent with our prior finding, which demonstrated larger extrafoveal avascular areas in the DCP in exudative AMD.[23]
- In DR, NPCs were frequently adjacent to microaneurysms, and the number of microaneurysms showed a strong correlation with NPC metrics. This is consistent with post-mortem histological observations by Cogan et al., who noted the localization of microaneurysms near zones of acellular, occluded capillaries in the early stages of DR.[20,45] They speculated that capillary occlusion in these zones plays a critical role in the development and orientation of microaneurysms, accounting for the clustered and circular arrangement of lesions observed ophthalmoscopically in DR.
- Interestingly, retinal fluid did not exhibit a significant relationship with NPC metrics in either disease, suggesting that NPCs may be more closely related to structural and vascular changes rather than fluid accumulation.

Despite these findings, there are several limitations to this study. First, multiple repeated scans are required for registration and averaging to improve scan contrast due to the low signal-to-noise ratio of a single OCT volume. This process is essential for extracting structural capillaries from OCT and identifying NPCs using co-registered OCTA. However, acquiring multiple repeated scans prolongs the total acquisition time, increasing the likelihood of image artifacts caused by eye movements and introducing additional challenges for clinical imaging workflow. With the rapid advancements in OCT technology, we anticipate this limitation will be mitigated soon. For instance, ultrahigh-resolution OCT[46] will reduce the number of scans needed, and higher-speed OCT will reduce the acquisition time. Second, the study cohort is relatively small. In future research, increased patient enrollment will allow for the collection of more data, enabling a more comprehensive study of NPCs.

In conclusion, a new method was developed to segment and quantify NPCs using co-registered OCT and OCTA modalities. Our study demonstrates the potential of advanced imaging techniques and deep learning models to transform the detection and characterization of NPCs in retinal diseases.

**Acknowledgment**

**Funding/Support:** This work was supported by the National Institute of Health (R01 EY027833, R01 EY035410, R01 EY024544, R01EY036429, R01 EY031394, T32 EY023211, UL1TR002369, P30 EY010572); the Malcolm M. Marquis, MD Endowed Fund for Innovation; an Unrestricted Departmental Funding Grant and Dr. H. James and Carole Free Catalyst Award from Research to Prevent Blindness (New York, NY), Edward N. & Della L. Thome Memorial Foundation Award, and the Bright Focus Foundation (G2020168, M20230081).

**Financial interests:** Yali Jia: Visionix/Optovue (P, R), Roche/Genentech (P, R, F), Ifocus Imaging (I), Optos (P), Boeringer Ingelheim (C), Kugler (R). Steven T. Bailey: Visionix/Optovue (F). Yukun Guo: Visonix/Optovue, Inc. (P), Genentech, Inc. (P, R). Jie Wang: Visionix/Optovue, Inc. (P), Genentech, Inc. (P, R).